# Latched Detection of Excited States in an Isolated Double Quantum Dot

A. C. Johnson, C. M. Marcus
*Department of Physics, Harvard University, Cambridge, Massachusetts 02138*
M. P. Hanson, A. C. Gossard
*Department of Materials, University of California, Santa Barbara, California 93106*

Pulsed electrostatic gating combined with capacitive charge sensing is used to perform excited state spectroscopy of an electrically isolated double-quantum-dot system. The tunneling rate of a single charge moving between the two dots is affected by the alignment of quantized energy levels; measured tunneling probabilities thereby reveal spectral features. Two pulse sequences are investigated, one of which, termed latched detection, allows measurement of a single tunneling event without repetition. Both provide excited-state spectroscopy without electrical contact to the double-dot system.

Electrically controllable discrete quantum states found in quantum dot systems are highly efficient laboratories for the study of quantum coherence [1], as well as a potential basis for quantum computation [2]. Measuring the energy spectrum and dynamics in quantum dots necessarily involves coupling to a macroscopic measurement apparatus, which in turn may act to reduce coherence [3]. Excited state spectroscopy of single [4] and double [5, 6] quantum dots has typically been performed using nonlinear transport, requiring direct (dc) coupling of the device of interest to electron reservoirs [6]. This coupling perturbs the quantum states and may increase decoherence and heat the device, particularly at the large biases needed for spectroscopy far from the Fermi surface.

An alternative approach that we investigate in this Letter is to use capacitive charge sensing [7, 8] combined with pulsed gate voltages that provide an excitation window [9]. Charge sensing has recently been used to probe excited-state spectra in a few-electron quantum dot configured to remain coupled to one reservoir [10]. We investigate pulse/sense spectroscopy in an electrically isolated double quantum dot, where a *single charge*, moving between the two dots, is used to probe excited states. Local charge sensing from a quantum point contact (QPC) located near one of the dots provides readout.

The pulse/sense method operates as follows: A *reset pulse* on two gates simultaneously opens the coupling between the dots and "tilts" the potential, putting the excess charge on a selected dot (a micrograph of the device is shown in Fig. 1a). The reset pulse is then removed. With each dot separately in its ground state, but the double dot system now out of equilibrium, the excess charge is given a finite time to tunnel to the other dot (the *probe* time). The probability of tunneling depends sensitively on the alignment of ground and excited state levels in the two dots. Whether or not the charge tunnels during the probe time can be readily detected using a QPC sensor, either during the probe time or after.

Two gate sequences are investigated. In the first, a short reset pulse is followed by a relatively long probe interval during which a low tunneling rate gives a moderate total probability for tunneling. By cycling the reset/probe steps, the sensing QPC measures the average charge configuration, dominated by the probe step. This allows fine energy resolution, as the probe process is insensitive to both thermal effects and experimental difficulties associated with short pulses.

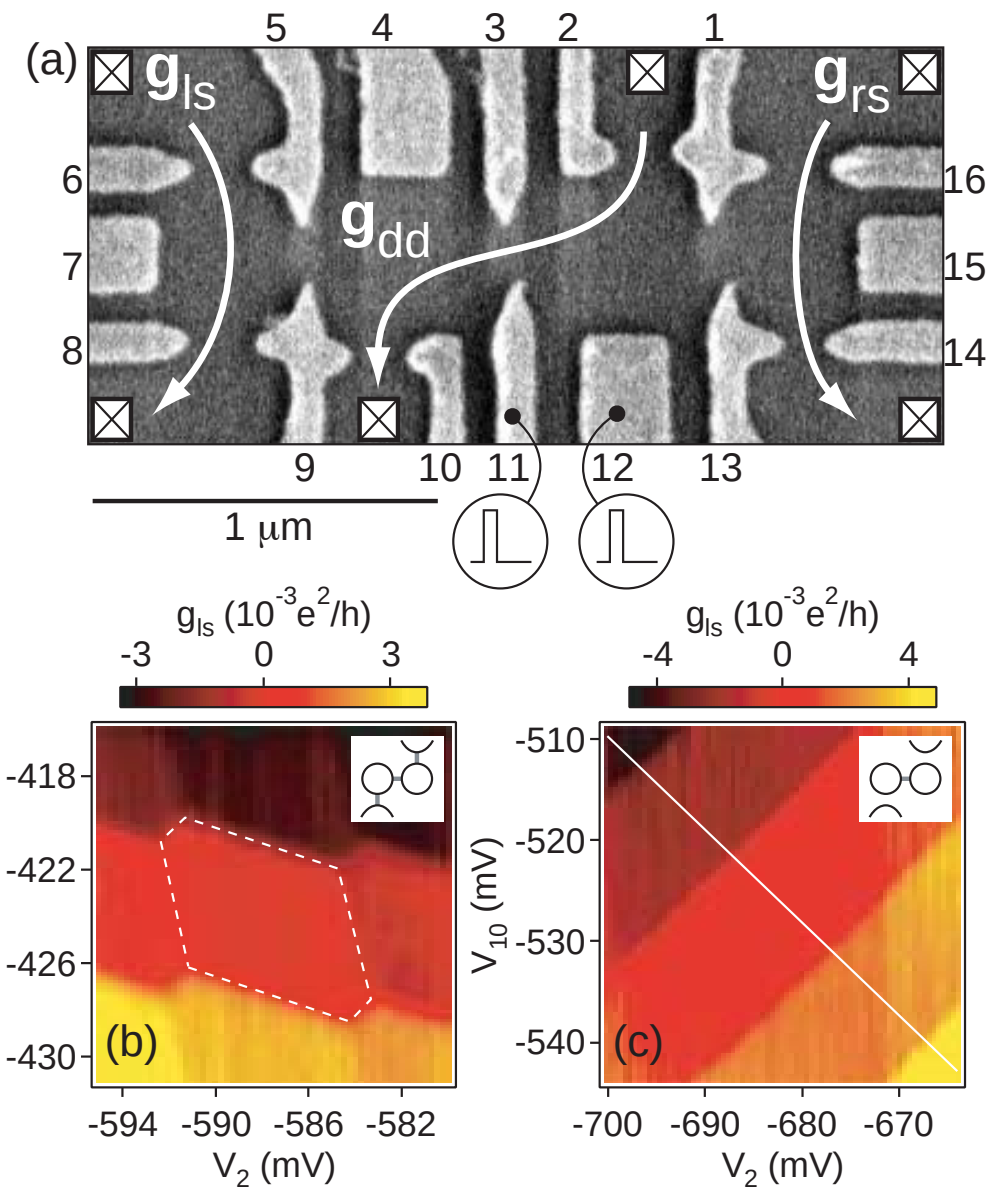

FIG. 1: (a) Scanning electron micrograph of a device identical in design to the one used in the experiment, consisting of a double quantum dot with charge sensor on either side. High-bandwidth coaxial lines are attached to gates 11 and 12; dc lines to the other gates. Sensors were configured as quantum point contacts by grounding gates 6, 7, 15, and 16. (b) Left sensor signal as a function of gate voltages $V_2$ and $V_1$ reveal hexagonal charge stability regions (one outlined with dashed lines) when the double dot is tunnel-coupled to leads. Predominantly horizontal features indicate that the left sensor has greater sensitivity to charge of left dot, controlled by $V_{10}$. (c) Sensing signal when the double dot is fully isolated from the leads. The only transitions are those transferring one electron between the two dots. In all pulsed-gate/sense experiments presented, the double dot is isolated (as in (c)) and $V_2$ and $V_{10}$ are swept simultaneously across the charge-transfer transitions along a diagonal (solid white line). A background plane is subtracted in (b) and (c) to compensate direct coupling between gates 2 and 10 to the charge sensor.

The second gate sequence uses two pulses: the first pulse resets the system, the second allows weak tunneling between dots. The second pulse is followed by an arbitrarily long interval (microseconds to hours, in principle) when the barrier between dots is closed, so that each dot is completely isolated, with fixed charge. This sequence we term *latched detection*, because the measurement occurs after all barrier have been closed, i.e., the double-dot system is latched into a final state, making all charge dynamics separated in time from the measurement. If needed, the measurement circuit could be turned off altogether during the probe pulse, though here we make the measurement time much longer than the probe time and measure weakly, so that the total back-action of the measurement on the system during the probe pulse is negligible. While we only apply latched detection to the problem of excited-state spectroscopy, we emphasize that its usefulness is much more general.

The device (Fig. 1(a)), defined by e-beam patterned Cr-Au depletion gates on a $GaAs/Al_{0.3}Ga_{0.7}As$ heterostructure grown by MBE, comprises two tunnel-coupled quantum dots of lithographic area 0.25 $\mu m^2$ each and two independent charge-sensing channels, one beside each of the central dots. The two-dimensional electron gas lies 100 nm below the surface, with bulk density $2 \times 10^{11}$ $cm^{-2}$ and mobility $2 \times 10^5$ $cm^2/Vs$. Each dot contains ~ 150 electrons and has a single-particle level spacing $\Delta$ ~ 100 μeV (estimated from effective device area) and charging energy $E_c$ ~ 600-700 μeV. Measurements were carried out in a dilution refrigerator with base electron temperature of ~100 mK. Left and right sensor conductances $g_{ls}$ and $g_{rs}$ were measured using lock-in amplifiers with 2 nA current biases at 137 and 187 Hz; the double-dot conductance $g_{dd}$ was measured using a third lock-in amplifier with a 5 μV voltage bias at 87 Hz, although during the pulse/sense measurements the double dot was fully isolated and the $g_{dd}$ and $g_{rs}$ circuits grounded. Throughout the experiment, the charge sensors were configured as QPCs by grounding gates 6, 7, 15, and 16, and were well isolated from the double dot by strongly depleting gates 5, 9, 13, and 1.

Figure 1(b) shows the left sensor signal as a function of gate voltages $V_2$ and $V_{10}$ when the device was tunnel-coupled to both source and drain leads. Here and in subsequent plots, a plane has been subtracted to level the central plateau to compensate for capacitive coupling between the gates and the sensor. In this regime, a honeycomb pattern characteristic of double-dot transport [6] is seen as a set of hexagonal plateaus in the left sensor conductance, with horizontally oriented steps of ~ $3 \times 10^{-3}$ $e^2/h$ corresponding to changes in the number of electrons in the left dot, controlled by $V_{10}$, and smaller vertically oriented steps marking changes in the right dot, controlled by $V_2$. Steps at the short segments at the upper left and lower right of each hexagon reflect movement of an electron from one dot to the other, with total number fixed. Resonant transport through the double dot, $g_{dd}$, in this weakly coupled regime occurs only at the honeycomb vertices [8].

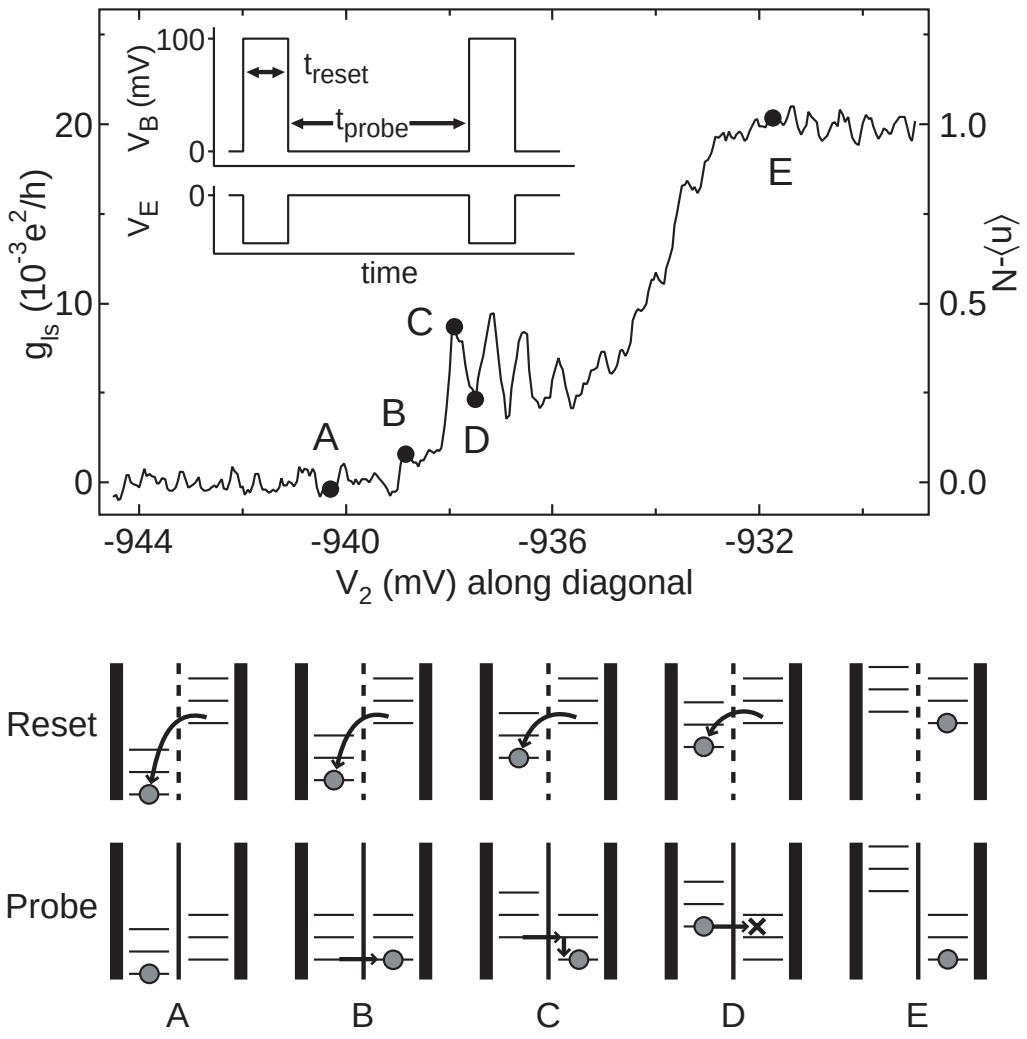

FIG. 2 Single-pulse technique. Time-averaged conductance of left sensor as a function of $V_2$ along diagonal (see Fig. 1(c)) with pulses applied. Inset: Pulses on gates 11 and 12 (applied in a linear combination parameterized by $V_B$ and $V_E$, controlling interdot barrier and relative energy, see text), followed by a long interval during which an electron may or may not tunnel. A linear fit to the left plateau is subtracted. Right axis shows the average excess occupation $\langle n \rangle - N$ of the right dot. Points A through E mark features used to infer the excited state spectrum, with schematic interpretations shown below the graph.

As the double dot becomes more isolated from the leads by making $V_2$ and $V_{10}$ more negative, the honeycomb sensing pattern persists long after $g_{dd}$ has become immeasurably small. However, upon isolating the double-dot system still further, the landscape changes dramatically as the tunneling time between the double dot and the leads diverges. This is the case in Fig. 1(c), where transitions appear in the sensor signals as diagonal lines, corresponding to lines of constant energy difference between the dots. An increase in $g_{ls}$ corresponds to an electron moving *away* from the left sensor, or an increase in the number of electrons in the right dot. Pulse/sense measurements were carried out in this isolated configuration, with the energy difference between the dots controlled by simultaneously varying $V_2$ and $V_{10}$ along diagonals (shown, for example, by the white line in Fig. 1(c)), and controlling the tunnel barrier between the dots with gate voltage $V_3$.

Fast control of the energy difference and barrier height was achieved using two synchronized Agilent 33250 arbitrary waveform generators, with rise times of ~5 ns, connected to gates 11 and 12 via semirigid coaxial lines and low-temperature bias tees. To compensate the slight cross-coupling of gates 11 and 12, the pulse generators produced linear combinations of pulses, denoted $V_B$ (affecting the tunnel, mainly $V_{11}$) and $V_E$ (affecting the energy difference, mainly $V_{12}$).

The single-pulse/probe sequence is shown schematically in the inset of Fig. 2. Square pulses of length $t_{reset}$ = 100 ns are simultaneously applied to gates 11 and 12, once every 20 μs, with $V_B$ = 100 mV to open the tunnel barrier significantly while the pulse energy shift $V_E$ is varied. The double-dot system will relax to its overall ground state during the reset pulse as long as $t_{reset}$ is much longer than the elastic and inelastic tunneling times $\tau_{el}$ and $\tau_{in}$ while the tunnel barrier is pulsed open, and is also longer than the energy relaxation time within the dots. During the remaining 19.9 μs, the tunnel barrier is nearly closed, such that $\tau_{el} < t_{probe} < \tau_{in}$, and the energy levels are returned to their values before the pulse. Under these conditions, if elastic tunneling is allowed, the electron will likely tunnel and then have ample time to relax to its ground state. If elastic tunneling is forbidden, however, the electron will likely remain where it was put by the reset pulse.

Figure 2 shows the sensing signal during a simultaneous ("diagonal", see Fig. 1c) sweep of $V_2$ and $V_{10}$, with five points (A to E) labeled to highlight key features. The data were taken with $V_E$ negative, so that the reset pulse tilts the ground state toward the left dot. At point A, the tilt of the double dot during the probe state favors the excess charge remaining in the left dot, and a flat sensing signal corresponding to a time-averaged right-dot occupation $\langle n \rangle = N$ is observed. At the opposite extreme (E), the energy shift during both the reset and probe states favors the excess charge occupying the right dot, which gives again a flat sensing signal, now corresponding to $\langle n \rangle = N + 1$. At point B, the ground states of the dots are degenerate during the probe time (except for small tunnel splitting). This degeneracy appears in the data as a small peak in right-dot occupation, often barely visible above the noise, presumably because tunneling in this case is reversible; the electron does not relax once it enters the right dot, and so it is free to tunnel back to the left dot. At point C there is a much larger peak. Here either an excited state in the right dot aligns with the ground state in the left or a hole excited state in the left dot aligns with the ground state in the right. When the electron tunnels the system will subsequently relax, trapping the electron on the right dot. Finally at D, no excited states exist to match the initial configuration and allow elastic tunneling, so there is a dip in the right-dot occupation.

Figure 3(a) shows the sensing signal measured throughout the pulse/dc-energy-shift plane, with prominent diagonal steps marking the ground state transitions during the reset pulse, and fine bands extending horizontally from each step, reflecting excited state transitions available during the probe time. Figure 3(b) presents the same data differentiated with respect to $V_2$ and smoothed along both axes. Here, steps in the raw signal appear as positive ridges, and excited state peaks become bipolar. Now the comparison with single-dot transport spectroscopy is much more evident: The pulse $V_E$ opens an energy window, and as the window expands new excited states become energetically accessible and emerge from the ground state feature. Features to the left (negative $V_E$) correspond to electron excited states of the right dot or hole excited states of the left; features to the right mark electron excited states of the left or hole excited states of the right.

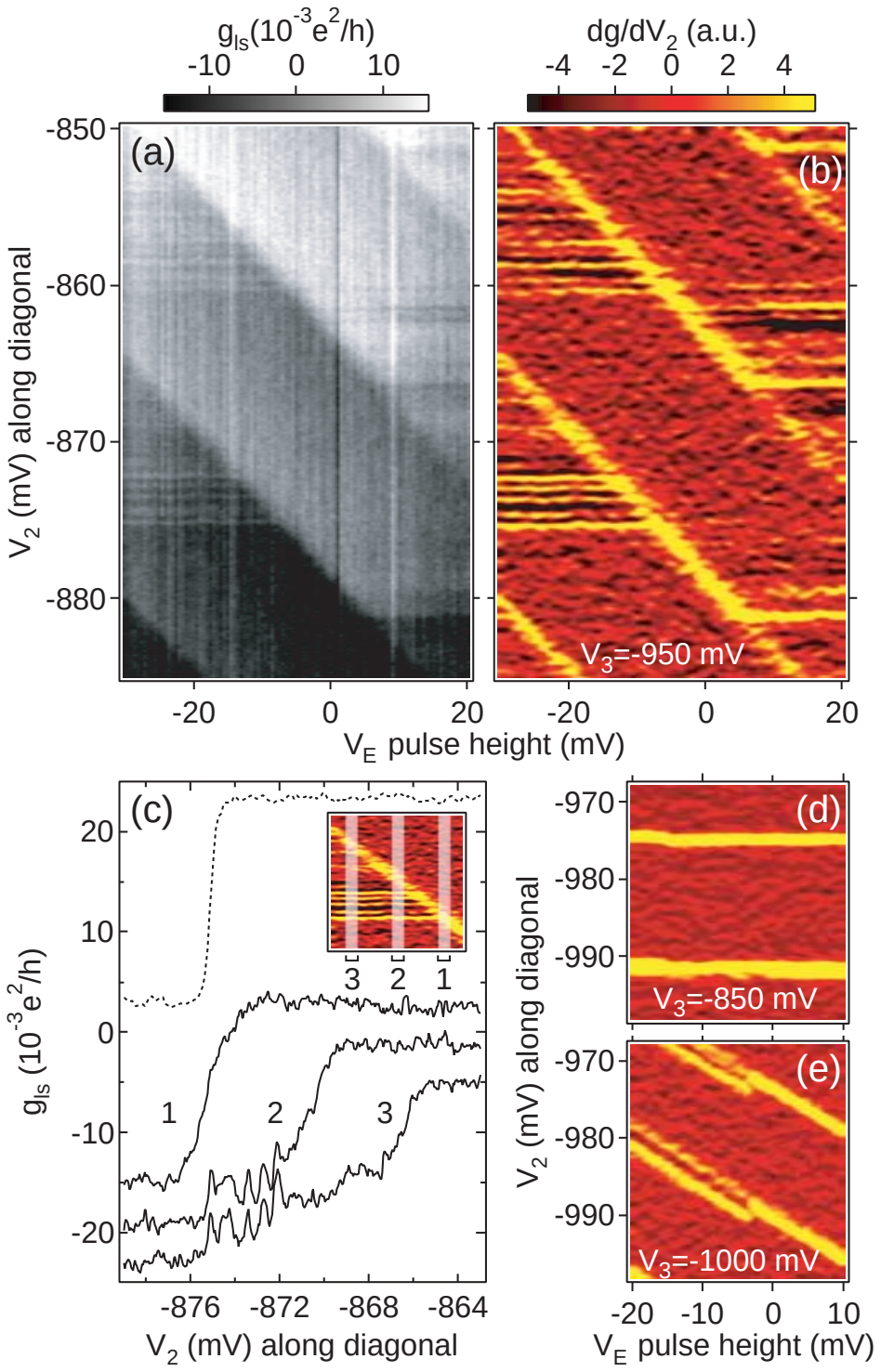

FIG. 3 (a) Left sensor conductance, and (b) its smoothed derivative with respect to $V_2$, as functions of $V_2$ and $V_E$. Horizontal excited state lines emerge from a diagonal ground state feature as the energy window $V_E$ is increased. Vertical stripes in (a) result from drift of the sensor. (c) Slices of conductance (as in Fig. 2) as a function of $V_2$, averaged over different $V_E$ ranges (see inset), offset vertically for clarity. Dashed curve shows a transition measured with pulses off. (d,e) Sensor conductance derivative as in (b) measured with the tunnel barrier either more open (d) or more closed (e) such that the probe or the reset configurations respectively dominate the measurement. Color scales in (d) and (e) are the same as in (b).

Figure 3(c) shows slices of sensor conductance from Fig. 3(a) at three different $V_E$ pulse heights, illustrating the expansion of the energy window while the positions of the emerging excited states remain fixed. The dashed curve shows the transition measured with no pulses applied. From the smooth, narrow shape of the no-pulse transition, and its consistency from transition to transition, we conclude that its width is dominated by temperature broadening [8]. Assuming the electron temperature in the dot does not depend significantly on the coupling to leads, we associate with this broadening a temperature of ~100 mK (calibrated in the tunneling regime [8]), which gives a lever arm $\delta V_2/\delta E = (10.5 \pm 1)/e$ relating changes in $V_2$ (along a diagonal with $V_{10}$) to changes in the energy of levels in the left dot relative to the right. Taking the spacing between transitions to be the sum of charging energies of the two dots, we find $E_c = 700 \pm 70$ μeV. The measured excited-state gate-voltage spacing of ~ 0.75 mV gives an excited level spacing of ~70 μeV, comparable to the Δ ~ 100 μeV estimate from dot area. The slightly lower measured value may be evidence of the sensitivity of the sensing signal to both electron and hole ex-cited states, giving overlapping spectra. As the energy window is increased, excited-state-to-excited-state transitions become available, further complicating the observed spectra. This may explain the blurring observed for $V_2 > -872$ mV in curve 3.

Figure 3(c) shows that the transitions from $\langle n \rangle = N$ to $N + 1$ for the pulse curves (solid) are clearly broader than for the no-pulse curve (dashed). Broadening beyond temperature is presumably due both to averaging traces with different $V_E$ as well as effects of overshoot and settling of the pulse. However, because the probe time is long and insensitive to pulse properties, the excited state peaks are not affected by these sources of broadening. In principle, the excited state peaks are also immune to thermal broadening, their widths limited only by intrinsic decay rates, although all peaks shown here exhibit full widths at half maximum of at least 3.5 $kT$ (0.3 mV in this case), possibly due to gate noise in the charge sensor.

Figures 3(d) and (e) repeat Fig. 3(b) but with different values of tunnel-barrier gate voltage $V_3$, illustrating the effects of opening or closing the tunnel barrier beyond the regime of excited-state spectroscopy. Changing $V_3$ affects the tunnel barrier between dots both during and after the pulse, although the system is most sensitive to the tunnel rate during the probe time. Increasing tunneling by making $V_3$ less negative by 100 mV (Fig. 3d) yields single, horizontal features in $dg/dV_2$, as if no pulse were applied at all. This implies that $\tau_{in} << t_{probe}$ so the system quickly finds its ground state during the probe time regardless of relative energy levels, making the pulse irrelevant to a time-averaged measurement. Reducing the tunneling rate by making $V_3$ more negative by 50 mV (Fig. 3e) results in diagonal features, indicating that dynamics during the pulse dominate behavior. In the right half of this plot there is a single transition, implying that the system finds its ground state while the pulse is on, then the barrier is closed such that $\tau_{el}$, $\tau_{in} >> t_{probe}$ and no further tunneling is permitted.

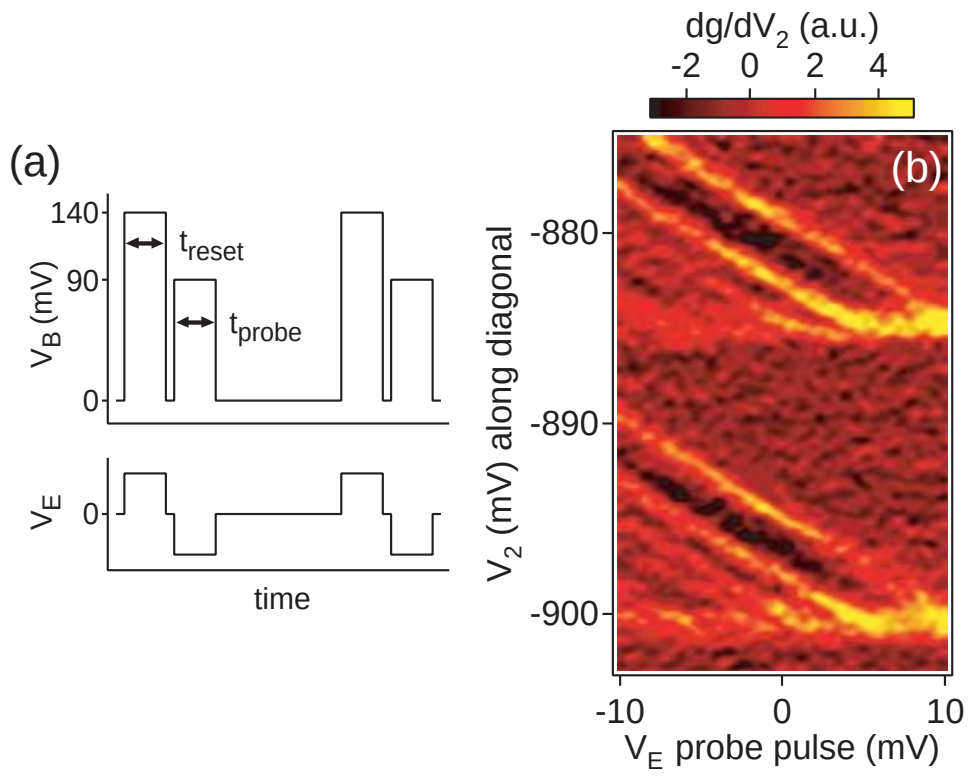

FIG. 4 (a) Two-pulse technique, shown schematically, including a reset and a probe pulse, followed by an arbitrarily long measurement time during which no tunneling is allowed. (b) Conductance derivative with respect to $V_2$, shows excited states appearing now as diagonal lines because the probe $V_E$ is being varied rather than the reset $V_E$.

On the left there are two diagonal features, implying that an excited state is populated at the start of the reset pulse. This effect is not understood at present, nor is it specific to the too-closed-barrier regime; it is also seen occasionally in conjunction with the understood horizontal excited state features.

We now turn to the second pulse/sense method described above, latched detection, which uses two pulses on each gate as shown in Fig. 4(a). Figure 4(b) shows the derivative of the sensor signal measured in this configuration as a function of $V_2$ and the $V_E$ probe pulse height, using $t_{reset} = t_{probe} = 20$ ns and $V_B = 140$ mV for the reset pulse and 90 mV for the probe pulse. Here, because we vary the probe properties rather than the reset properties as in Fig. 3, excited states appear as diagonal features and the reset ground state is the horizontal feature. Excited states measured this way do not have the same immunity to pulse shaping enjoyed by excited states measured with the single-pulse method, and as a result, the data in Fig. 4 are blurred relative to Fig. 3. This diminished resolution is not fundamental, and can be reduced with more accurate pulse shaping.

We thank M. J. Biercuk and A. Yacoby for useful discussions and K. Crockett for experimental contributions. This work was supported in part by DARPA QuIST, Harvard NSF-NSEC, and iQuest at UCSB.